# Symmetry Breaking and Topological Defect Formation in Ion Coulomb Crystals

K. Pyka[1], J. Keller[1], H.L. Partner[1], R. Nigmatullin[2,3], T. Burgermeister[1], D.M. Meier[1], K. Kuhlmann[1], A. Retzker[4], M.B. Plenio[2,3,5], W.H. Zurek[6], A. del Campo[6,7] & T.E. Mehlstäubler[1]

[1]*Physikalisch-Technische Bundesanstalt, Bundesallee 100, 38116 Braunschweig, Germany*
[2]*Institute for Theoretical Physics, Albert Einstein Allee 11, Ulm University, 89069 Ulm, Germany*
[3]*Department of Physics, Imperial College London, Prince Consort Road, London, SW7 2AZ, United Kingdom*
[4]*Racah Institute of Physics, The Hebrew University of Jerusalem, Jerusalem 91904, Givat Ram, Israel*
[5]*Center for Integrated Quantum Science and Technology, Ulm University*
[6]*Theoretical Division, Los Alamos National Laboratory, Los Alamos, NM, USA*
[7]*Center for Nonlinear Studies, Los Alamos National Laboratory, Los Alamos, NM, USA*

Symmetry breaking phase transitions play an important role in nature. When a system traverses such a transition at a finite rate, its causally disconnected regions choose the new broken symmetry state independently. Where such local choices are incompatible, defects will form with densities predicted to follow a power law scaling in the rate of the transition. The importance of this Kibble-Zurek mechanism (KZM) ranges from cosmology to condensed matter [1-4]. In previous tests in homogeneous systems, defect formation was seen, but weak dependence on the transition rate and limited control of external parameters so far prevented tests of KZM scaling. As recently predicted [5-9], in inhomogeneous systems propagation of the critical front enhances the role of causality and steepens scaling of defect density with the transition rate. We use ion Coulomb crystals in a harmonic trap to demonstrate, for the first time, scaling of the number of topological defects with the transition rate – the central prediction of KZM - in a well-controlled environment.

Ion chains confined in harmonic traps were investigated as a promising platform for quantum information processing [10,11]. We use them to emulate the dynamics of symmetry breaking: As the triaxial potential confining an ion chain is weakened in the transverse direction, the chain buckles, and the ions form a "zigzag" (see Fig. 1). This breaking of the (axial) symmetry is a second order phase transition described by the Landau-Ginzburg theory [12]. It can lead to formation of topological defects (places where e.g. a "zig" is followed by another "zig" rather than a "zag"), as highlighted in Fig. 1 in red circles. Depending on the speed at which the transverse potential is quenched, the system chooses a final configuration of a pure "zigzag", or with kinks dividing the two different orientations. The appearance of such kinks, showing soliton-like behavior [20], is a result of the incompatible local choices of symmetry breaking.

Phase transitions have traditionally been studied in equilibrium, with a focus on the critical region, where critical slowing down and critical opalescence - i.e. divergence of relaxation time and correlation length near the critical point - allow classification of second order phase transitions into universality classes that are independent of the micro-physics of the system. However, critical slowing down implies that second order phase transitions inevitably depart from equilibrium near the critical point, where the broken symmetry is chosen. That choice is made locally, within regions (domains) that can coordinate symmetry breaking. Cosmology provides well-known examples: as noted by Kibble [1,2], relativistic causality limits the size of domains where consensus about symmetry breaking can be reached. As a result of disparate choices of broken symmetry, topological defects (monopoles, cosmic strings, and domain walls) can form, with dramatic consequences for the ensuing cosmological evolution.

In laboratory phase transitions, relativistic causality is not a useful constraint - it does not lead to predictions of defect density. Here, the speed of light is replaced by the velocity of sound and to estimate the domain size one can instead appeal to critical slowing down [3,4]: near second order phase transitions, the relaxation time $\tau$ (characterizing reflexes of the system) and the healing length $\xi$ (that sets the scale on which the order parameter returns to its equilibrium value) diverge as $\tau = \tau_0/|\epsilon|^{\nu z}$ and $\xi = \xi_0/|\epsilon|^\nu$.

Above, $\tau_0$ and $\xi_0$ depend on microphysics, while the critical exponents $\nu$ and $z$ define the universality class of the transition. The distance from the critical value is given by the dimensionless parameter $\epsilon$ (e.g., $\epsilon = (\lambda_c - \lambda)/\lambda_c$ where $\lambda_c$ is the critical value of the control parameter $\lambda$ at which the phase transition occurs). It is the change of $\epsilon$ with time that takes the system from the symmetric ($\epsilon < 0$) to the symmetry broken ($\epsilon > 0$) phase. In what follows we take $\epsilon = t/\tau_Q$. The ratio of $\xi$ and $\tau$ yields the speed of sound $v = (\xi_0/\tau_0)|\epsilon|^{\nu(z-1)}$. This is the speed with which the information about the choice of broken symmetry can propagate. It (along with the resulting sonic horizon) plays a key role in the formation of defects.

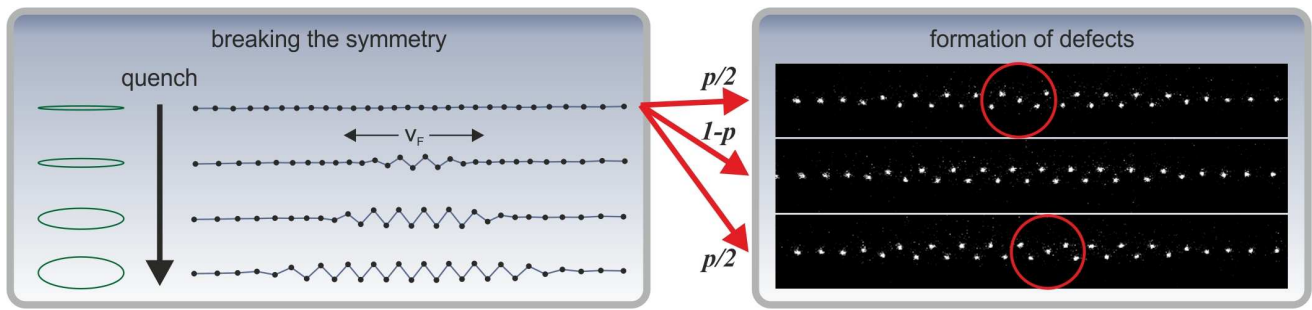

**Figure 1| Phase transition in an inhomogeneous system leading to kink formation.** Left: An ion chain buckles in the center and the "zigzag" region spreads at a finite velocity $v_F$ during a quench of the transverse confinement (illustrated by the green ellipses). Right: Examples of the configuration after a quench: normal "zigzag" (center) and kink and anti-kink (top and bottom). The system chooses the final configuration with a certain probability, depending on the speed with which the phase transition is crossed.

The size of domains that can coordinate symmetry breaking is set at the instant $\hat{t}$ when, because of critical slowing down, the state of the system ceases to follow changes imposed by the time-dependent control parameter $\epsilon$. This happens when the relaxation time matches the timescale of the changes imposed on the system, which leads to the following equation for $\hat{t}$: $\tau[\epsilon(\hat{t})] = \epsilon/\dot{\epsilon}$. The solution $\hat{t} = (\tau_0 \tau_Q)^{1/(1+\nu z)}$ determines the size of the sonic horizon $\hat{\xi} = \xi_0 (\tau_Q/\tau_0)^{\nu/(1+\nu z)}$ and the typical separation between topological defects.

In [6], Coulomb crystals trapped in a harmonic axially confining potential were proposed to measure the scaling of defect formation to test KZM. In such crystals, the charge density along the ion chain is larger in the center and reduces towards the edges. Within the local density approximation, this inhomogeneity is inherited by the critical control parameter $\lambda_c$. In our case, this is the transverse secular frequency describing the radial confinement, whose spatial dependence is given by $\nu_{t,c}(x) \simeq 2\omega(x)$. Here $\omega(x) = \left[\frac{Q^2}{m\, a(x)^3}\right]^{1/2}$ is the characteristic frequency associated with the axial Coulomb coupling given in terms of the mass $m$ and charge $Q$ of the ion, and the axial inter-ion spacing $a(x)$ which reaches its minimum value at the origin.

Laser cooling provides a dissipative force acting on the ions. In the underdamped regime accessible in our experiment, the characteristic response time is fixed by the inverse of $\omega(0)$ and independent of the friction parameter $\eta$. As a result, the system exhibits mean-field critical exponents $\nu = 1/2$ and $z = 1$ [6].

We consider a quench of the transverse secular frequency of amplitude $\delta$ in a time scale $\tau_Q$, i.e., $\nu_t = \nu_{t,c}(0) - \delta\, t/\tau_Q$. During the quench, the transition is first crossed at the center of the chain. The velocity of the front $\nu_F$ with which the critical point propagates depends on the quench rate and varies along the chain. The dynamics of an inhomogeneous phase transition is governed by the interplay of $\nu_F$ and the sound velocity $\nu = \xi_0/\tau_0 = \omega(0) a(0)$. Under a fast quench, the inequality $\nu_F > \nu$ holds everywhere in the system, and the paradigmatic Kibble-Zurek mechanism for homogeneous systems (HKZM) is recovered. The sharpness of this inequality was studied in detail (see Supplementary Figure 5). For mean field-critical exponents in the underdamped regime the density of kinks d scales with the quench rate as $d \propto \tau_Q^{-1/3}$ in agreement with the HKZM, as seen in Fig. 2. At lower quench rates, $\nu_F > \nu$ is only satisfied in a given fraction of the system $\hat{x}$, which is smaller than unity and depends itself on the quench rate. This is the regime where defects can be produced by the Inhomogeneous Kibble-Zurek mechanism (IKZM) [5,6]. Outside $\hat{x}$, the critical front is slower than $\nu$, and defects will not form [13,14]. As a result, the total number of kinks is substantially reduced and exhibits a more pronounced dependence on the quench rate, i.e. $d \sim \hat{x}/\hat{\xi} \sim (\tau_0/\tau_Q)^{(1+2\nu)/(1+\nu z)} \propto \tau_Q^{-4/3}$, see Fig. 2. For parameters such that typical runs lead to one or no defects, a doubling of the Kibble-Zurek scaling has been reported in homogeneous systems in a series of works [15-17], and verified experimentally [18]. The probability $p_{1kink}$ to observe a single kink in an inhomogeneous system obeys a doubling of the IKZM scaling as well (DIKZM)

(Eq. 1) $\quad p_{1kink} \propto \frac{L^4}{\nu_{t,c}(0)^4 \nu^4} \left(\frac{\nu_{t,c}(0)\,\delta}{\tau_Q}\right)^{\frac{2(1+2\nu)}{(1+\nu z)}} \propto \tau_Q^{-8/3}$,

where $L$ is the half-length of the chain. This dependence can be observed in the lower left corner of Fig. 2. For details on IKZM and DIKZM see Supplementary Discussions III and IV. A remarkable consequence of the IKZM is the strong dependence of the density of kinks on the quench rate, which facilitates the test of universality in the dynamics across the phase transition, and the control of the number of kinks in the broken-symmetry phase.

In previous experiments, the scaling of defect formation in second order phase transitions has been studied in homogeneous regimes [17,19]. For a more complete overview on KZM experiments see Supplementary Discussion V. The main difficulty with observing the KZM scaling is the power law dependence of defect densities that requires varying the control parameter over a large range to obtain a high sensitivity to the scaling coefficient. The excellent experimental control over trapped ions allows a flexible span of several orders of magnitude in the quench rate of the trapping potential, making them an ideal system for such tests. Moreover, linear traps offer the possibility of measuring in the inhomogeneous regime where the sensitivity is improved due to the enhanced scaling. In the measurements presented here, we verify the prediction for the DIKZM regime where at most one kink per quench is created and the scaling exhibits an 8-fold enhancement with respect to the homogeneous scenario. The ability to measure in the underdamped regime means that we do not have to make assumptions regarding the values of the critical exponents, since they can be derived using the Ginzburg-Landau model ($\nu = 1/2$ and $z = 1$). Finally, ion traps offer great scalability that will in principle allow one to probe other regimes of universal dynamics as well.

In Coulomb crystals, two types of metastable defect configurations that are robust against thermal fluctuations have been identified in numerical simulations, i.e. odd and extended kink solitons [6,20], and kink formation during the process of crystallization induced by laser cooling was observed experimentally [21]. To create kinks in our system [22] via a controlled symmetry breaking, we induce an ellipticity of 30% in the radial potential to confine the ions to a plane while we sweep the transverse frequency $\nu_t$ linearly over the critical value. The structure of the Coulomb crystals is detected and analyzed by imaging the fluorescence of the ions onto a camera. Figure 3 shows the experimentally obtained stable configurations of both types of kinks.

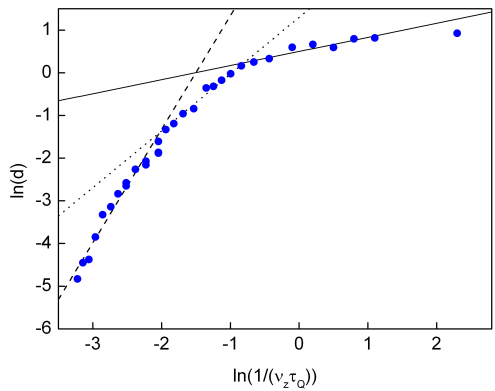

**Figure 2| Three regimes of kink creation**. Numerical simulation of the kink density as a function of the quench time $\tau_Q$ using our experimental parameters. Three regimes are identified (HKZM, IKZM and DIKZM) and indicated by the plotted lines with slopes of 1/3, 4/3 and 8/3, respectively.

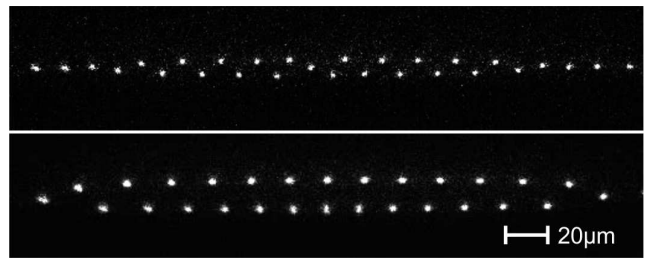

**Figure 3| Pictures of experimentally realised kink configurations. Top:** Odd kink, dominant for shallow quenches ($\nu_t/2\pi$ = 500 kHz to 204 kHz). **Bottom:** Extended kink, dominant for deep quenches in radial frequency from $\nu_t/2\pi$ = 500 kHz to 140 kHz. Exposure times are 50 ms and 200 ms, respectively. The Coulomb crystals consist of 29±2 ions and the phase transition sets in at $\nu_{t,c}/2\pi \approx 275$ kHz. For the statistics measurements, we use the extended kink configuration to preserve and stabilize the created localized odd kinks.

For quench times $t = 2\tau_Q$ on the order of the oscillation period $T = 1/\nu_z$ of the axial secular motion of the ions, kink formation in the 2D "zigzag" structure is observed as a break in the periodic order of the ions' positions. Depending on the final radial inter-ion spacing odd or extended kink configurations are stable. The statistics described in this work are based on measurements of extended kinks, since the larger energy barrier for a defect switching sites, i.e. the Peierls-Nabarro potential [23], in this configuration gives an inherently higher stability compared to odd kinks. As extended kinks are evolving from initially created odd kinks during the radial quench and no more than one defect per quench is created in our experimental regime, both final configurations show the same statistics, which allows us to use extended kinks to test KZM scaling. In the Supplementary Information, we provide videos of kink creation for both types.

To obtain the statistics of kink formation, the radial quench cycle is repeated up to 2000 times for each quench time and the density of defects $d$ is defined as the average number of kinks per quench cycle. We measure the kink density as a function of the quench time $\tau_Q$ at a fixed axial frequency $\nu_z$.

The kink densities obtained in our experiment range from 0.01 up to 0.24, i.e. from one kink for every 100 quenches up to one kink for every 4-5 quenches, respectively. Background gas collisions that heat up the Coulomb crystal followed by nonadiabatic re-crystallization due to the laser cooling lead to a finite rate of spontaneous kink formation. Because of the low laser cooling rate (see Full Methods for details), these are rare events in our case and can be neglected when compared to the number of defects created during the induced quenches of the trapping potential. We experimentally observed a rate of 1 spontaneous kink in 67 s of observation time in the "zigzag" configuration.

In finite linear systems losses can occur, not only when pairs of kinks annihilate but in our case also when kinks move out of the "zigzag" region. The Peierls-Nabarro potential builds up during the quench from the center and decreases to zero towards the outer linear parts of the crystal. The mobility of kinks is determined by the depth of this potential in comparison with their kinetic energy, which is gradually dissipated by the phononic coupling to the laser cooled crystal, at a rate quantified by the friction parameter $\eta$. Once a kink is able to change sites, it experiences an outward acceleration due to the inhomogeneity of the potential, enhancing the losses.

We performed numerical simulations with our experimental parameters to investigate the dynamics that can lead to losses of kinks before they can be detected experimentally. As a result, we have identified two relevant mechanisms that affect different regimes of the quench rate. For very slow quenches, losses occur due to the fact that defects initially move faster than the phase transition front and can therefore escape into the linear region before the confining potential has built up sufficiently. For fast quenches, an enhanced amount of kinetic energy is introduced into the system that allows defects to leave the "zigzag" region even after the quench has completed. This leads to a significant amount of losses with a strong dependence on $\eta$.

To separate the effect of these losses from the density of kinks predicted by KZM, the time evolution of the number of defects for each quench cycle is determined in our simulations. We compare the absolute density of defects created during one cycle with the density of defects that remain after the ion motion has been damped out by the optical cooling forces (depending on the value of $\eta$, this is at 350 to 500 µs after the radial quench). After this time, a constant number of defects is observed in the simulations, defining the density of stable kinks that are detectable in our experiment. Their lifetime is only limited by background gas collisions and was experimentally determined to be 1.6 s (see Supplementary Discussion II).

Figure 4a shows the simulation results for the created and experimentally detectable kink densities for different values of $\eta$. There is a region with negligible losses from $\ln[1/(\nu_z\tau_Q)] \approx -2$ to $\approx -2.5$ where the experimental detection is suitable to directly observe the density of created kinks.

In Fig. 4b, the experimentally determined densities are shown, compared to simulated results for various values of $\eta$. A power law fit to the experimental data in the region from $\ln[1/(\nu_z\tau_Q)] = -1.9$ to $-2.6$ yields a scaling of $\sigma = 2.7 \pm 0.3$. This is in good agreement with the value of $\sigma = 2.63 \pm 0.13$ obtained in a fit to the numerical data for the amount of kinks created (filled symbols in Fig. 4a) from $\ln[1/(\nu_z\tau_Q)] = -2$ to $-3.2$. The matching amount of losses at fast quenches confirms the assumed friction coefficient in the simulations. An excellent repeatability of these measurements over several months was observed.

Another important aspect seen in Fig. 4a is the independence of the density of created kinks (filled symbols) from the friction coefficient $\eta$. This proves that the phase transition happens in the underdamped regime, in agreement with the choice of the critical exponent z = 1 [6]. For quench times $\ln[1/(v_z\tau_Q)] < -1.7$ the probability to produce more than one kink per ion chain becomes negligible (i.e. $1 \approx p_{1kink}$). This is the DIKZM regime, where one expects a scaling of $\sigma = 8/3$ for $p_{1kink}$ according to Eq. (1), which is in agreement with the observed values in both the experiment and the simulations.

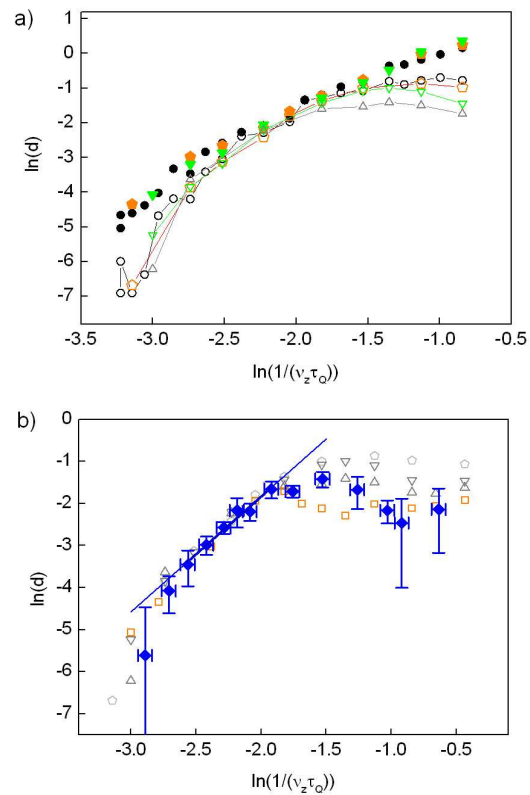

**Figure 4| Number of produced defects during radial quench. a,** Numerical simulations of the number of kinks created during a radial quench (filled symbols) as a function of the quench time $\tau_Q$ compared to number of kinks remaining at the time of detection (open symbols) for various values of the friction parameter $\eta$. The symbols correspond to $\eta = 0.34\ \nu_z$ (circles), $0.14\ \nu_z$ (pentagons), $0.09\ \nu_z$ (inverted triangles), $0.07\ \nu_z$ (triangles) with respect to the axial trap frequency $\nu_z/2\pi = 24.6$ kHz. The filled circles show the same data as Fig. 2. **b,** Experimentally measured kink density (filled symbols) compared to numerical simulations for $\eta = 0.14\ \nu_z$ (pentagons), $0.09\ \nu_z$ (inverted triangles), $0.07\ \nu_z$ (triangles), $0.06\ \nu_z$ (squares). The shown vertical errors include the standard deviation of measurements repeated over several days/months as well as the individual sample standard deviation assuming a binomial distribution. Horizontal errors are due to the uncertainty in the axial trapping frequency and a nonlinear distortion of the applied ramp.

The experimentally accessible regime could be extended by reducing losses at fast quench times. This could be achieved by enlarging the size of the crystal or by increasing the friction due to laser cooling (as seen in Fig. 4a), e.g. by using a lighter ion species. In this way, future experiments with Coulomb crystals may be able to measure the scaling in the other two regimes shown in Fig. 2. Since accessing the IKZM regime will require to increase the probability of creating more than one kink per chain, a stable detection of the more localized odd kinks becomes necessary to avoid long range interactions between defects. The homogeneous KZM scaling could be tested with ions trapped in homogeneous ring trap geometries [24].

In addition to demonstrating a well-controlled system for kink soliton creation and the first experimental test of KZM scaling in a causality enhanced regime, our results open up ways to explore the various properties of discrete non-linear stable excitations which have attracted considerable theoretical and experimental effort [25,26]. The often counter-intuitive non-linear physics of solitons, breathers and their thermal fluctuations becomes accessible in such systems [27]. In particular, our further research will explore kink dynamics, diffusion and interaction in the presented platform along with further numerical simulations, that have shown excellent quantitative agreement with our measurements. Assisted by laser cooling, thermal fluctuations could be reduced down to the quantum level, giving the unique ability to study the quantum properties of the defects [20] and to pursue the question of how KZM extends to quantum phase transitions [28,29].

**Acknowledgements** We thank B. Damski for reading of the manuscript, L. Yi for his contributions to the detection software, and K. Thirumalai for assistance in the lab. This work was supported by NSF PHY11-25915, the U.S Department of Energy through the LANL/LDRD Program and a LANL J. Robert Oppenheimer fellowship (A.d.C), the EU STREP PICC, the Alexander von Humboldt Foundation (M.B.P.), by EPSRC (R.N.) and by DFG through QUEST. A.d.C. and W.H.Z. are grateful to KITP for hospitality.

# Methods and additional information

## I.  EXPERIMENTAL DETAILS

**Atomic system and laser cooling.**

We laser cool $^{172}$Yb$^+$ ions on the $^2S_{1/2} - {}^2P_{1/2}$ transition at a wavelength of 370 nm to temperatures of a few mK, slightly above the Doppler cooling limit [30] of T = 0.5 mK for this transition. Two repump lasers at wavelengths of 935 nm and 638 nm deplete the populations of higher lying metastable states that are coupled to the cooling cycle via radiative decay or collisions. A single coil on top of the vacuum chamber produces a quantization field of about $B$ = 0.2 mT at the ion position to optimize the repumping efficiency of the 935 nm laser. A detailed description of the experimental setup can be found in [22].

Laser cooling the ions via a single travelling wave leads to a friction coefficient of the optical cooling force given by $\eta = \frac{2s}{(2+s)^2}\hbar k^2$ at $\Delta = -\Gamma/2$ [30], where $s$ is the saturation of the cooling transition, $k$ the wave vector of the cooling laser and $\Delta$ the detuning of the laser with respect to the resonance of the transition with linewidth $\Gamma$. The transverse beam profile is focused down to horizontal and vertical waist sizes $\omega_x$ = 8.8 mm and $\omega_y$ = 80 µm. For a laser power of 630 µW this results in an estimated experimental friction coefficient of $\eta_{exp}$ = 4×10$^{-21}$ kg s$^{-1}$ along a single axis. The simulations that have been carried out apply a friction parameter $\eta_{sim}$ to each axis and show the most agreement with the experiment for $\eta_{sim}$ = (2.5…3.0)×10$^{-21}$ kg s$^{-1}$.

**Ion trap and detection.**

The ion trap is a three dimensional segmented linear radio frequency (rf) Paul trap, which was designed for high precision spectroscopy on linear ion crystals. It offers full control to compensate stray fields in 3D and in particular, has low axial micromotion over a large range of several hundreds of micrometers [22]. The ions are trapped in a loading segment via photoionization and then shuttled to a spectroscopy segment, which is protected from the atomic beam. Avoiding contamination of the electrodes makes it possible to perform measurements with highly reproducible experimental parameters. The axial and radial secular frequencies in this segment were measured repeatedly, and a maximum deviation of $\sigma_z$ = 0.5 kHz and $\sigma_t$ = 1 kHz over several weeks was observed. For a linear crystal of 29 ± 2 ions with length $l$ ~ 300 µm the maximum axial rf field component along the crystal is as low as 500 V/m, whereas the radial field does not exceed 200 V/m, corresponding to a total micromotion amplitude of only $x_{mm}$ = 12 nm.

The ion crystal is imaged with a self-built lens system optimized for minimum aberrations onto an EMCCD camera with a magnification of M = 28 and an experimentally estimated resolution of about 1.5 µm over the full ion chain length of 300 µm. At the given axial trap frequency, about 30 ions can be imaged in the zigzag configuration.

**Modulation of rf voltage.**

The rf voltage driving the radial confinement is amplitude modulated by a control signal using an rf mixer. Its actual amplitude on the ion trap electrodes is monitored by a short antenna inside the helical resonator ($U_{mon}$) which impedance matches the amplified rf voltage $U_{rf}$ to the trap capacitance [22]. By observing parametric heating of the ions we can measure the radial as well as axial secular frequencies directly as a function of the trap voltages with a relative precision of 10$^{-3}$.

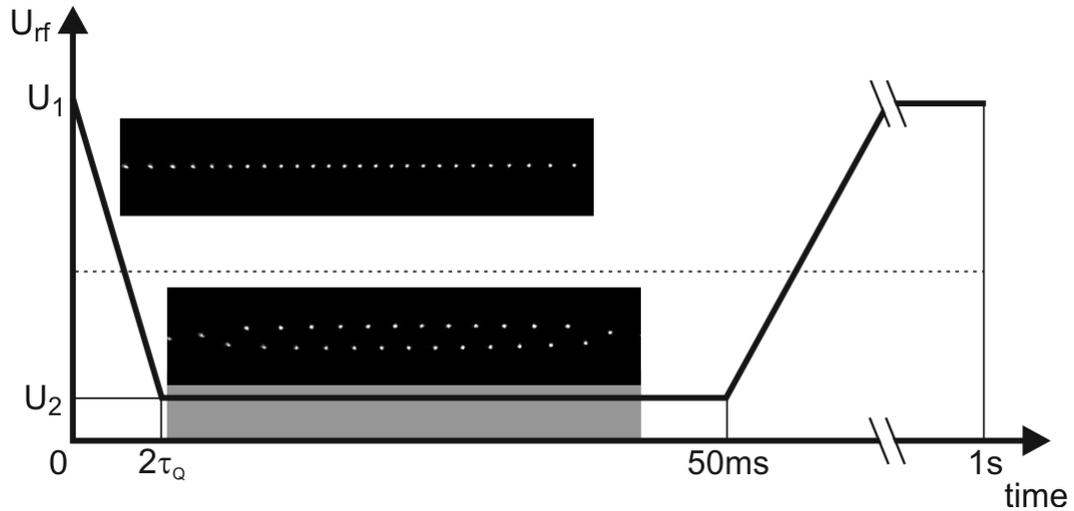

**Supplementary Figure 1 | Experimental sequence of the radial quench.** The radial trapping potential is steered by a linear ramp on the trapping voltage $U_{rf}$. The ramp of length $2\tau_Q$ starts at t = 0. Immediately after the quench an image of the Coulomb crystal is taken with an exposure time of 40 ms, then $\nu_t$ returns to its initial value. The ions are laser cooled and fluoresce throughout the whole sequence. After 1s the cycle is repeated.

To drive quenches of the radial confinement, a linear function, shown schematically in Supplementary Figure 1, is applied as a control signal to the rf mixer. Due to the characteristics of the mixer, the measured slope of the ramp deviates from the ideal slope and therefore is fitted manually to the monitored signal $U_{mon}$, see Supplementary Figure 2, to obtain the effective quench time used in the analysis shown in Fig. 4b (main text). The characteristic time constant of the resonant circuit of 7 µs limits the fastest possible quench time, but its effect on the ramps used in this work is negligible.

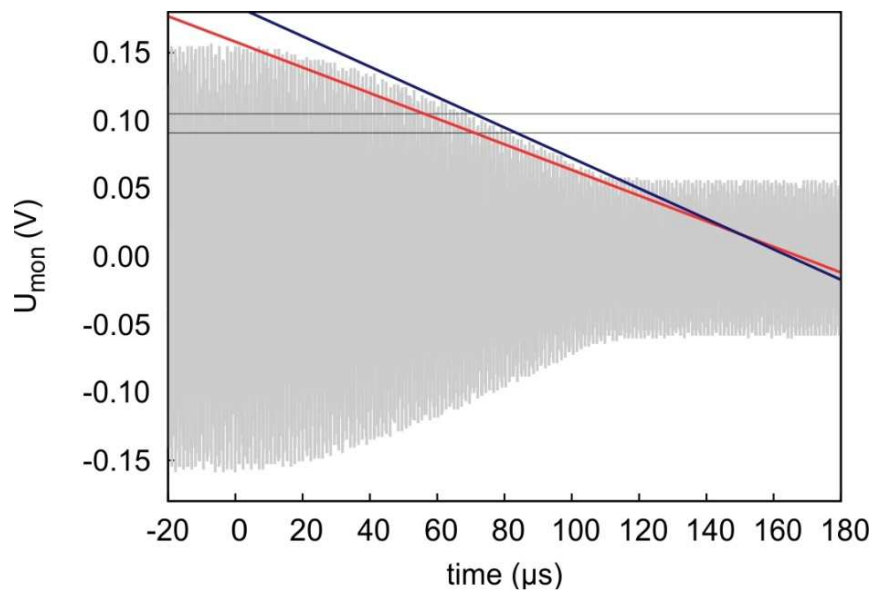

**Supplementary Figure 2 | Record of the monitored rf signal during the radial quench.** The linear control signal (red line) has a ramp time of 105 µs. The actual rf voltage ramp (light grey line) deviates from this due the nonlinearity of the rf mixer. Its slope is determined by a manual fit (blue line) and corresponds to an effective ramp time of 88.4 µs. Also shown is the region of the linear to zigzag transition (area between the horizontal grey lines) for $N = 29 \pm 2$ ions.

The ramp has also been tested for linearity in trap frequency by verifying that a linear fit to the ramp near the point of the linear-zigzag transition deviates by less than 2% from an ideal linear function between the frequencies measured at the extreme values of the ramp (see Supplementary Figure 3).

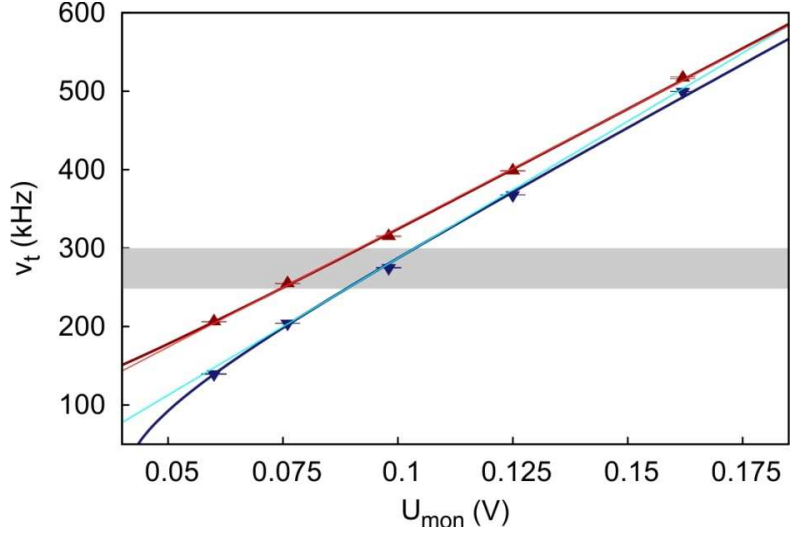

**Supplementary Figure 3 | Radial trap frequencies.** Values for both principal axes are measured by parametric heating of a single ion at different rf voltages. The bold lines show a fit using the analytical expression for the trap frequency $\nu_t = \frac{\Omega}{2}\sqrt{a + 0.5q^2}$. Here a and q are the Mathieu parameters and $\Omega$ is the frequency of the rf drive. The thin lines show a linear approximation of this function in the region of the linear to zigzag transition (grey bar). This linear approximation deviates by less than two percent from a linear function that directly connects the frequencies before and after the ramp. The width of the linear to zigzag transition region is due to variations in the number of loaded ions ($N = 29 \pm 2$).

**Error calculation and statistics.**

For each quench time, up to 2000 pictures are taken in several series of measurements over three months. Accounting for this, the kink density for each quench time is calculated as the mean of the densities from the individual series for that time weighted by the number of pictures per series. The first and biggest contribution to the error bars is the scatter of the individual series at a given quench time, calculated as sample standard deviation. The second and independent contribution accounts for statistical uncertainty of the data assuming a binomial distribution.

**Numerical simulations.**

For Doppler cooled ions the energy distribution of the ion chain obeys the Fokker-Planck distribution [31]. Hence, the dynamics of the *j*th ion can be described by the Langevin equation:

$$m\ddot{r}_j = -\eta\dot{r}_j + \nabla_j V_t(t) + \nabla_j V_c + \varepsilon_j(t), \quad (*)$$

where $V_t(t) = 1/2\sum_{i=1}^{N}(\nu_{t1}^2(t)x_i^2 + \nu_{t2}^2(t)y_i^2 + \nu_z^2 z_i^2)$ is the ponderomotive trapping potential, $V_c = e^2/(4\pi\epsilon_0)\sum_{i<j}^{N}|r_j - r_i|^{-1}$ is the Coulomb potential energy, $\eta$ is the friction coefficient due to laser cooling and $\varepsilon_j(t)$ is the stochastic force. The amplitude of the stochastic force, $\varepsilon_j(t)$, is related to the temperature $T$ and the friction coefficient $\eta$ by the fluctuation-dissipation relation $\langle\varepsilon_{\alpha j}(t)\varepsilon_{\beta k}(t')\rangle = 2\eta k_B T \delta_{\alpha\beta}\delta_{ij}\delta(t-t')$, with $\alpha, \beta = t_1, t_2, z$. Collisions with background particles are not included in this model.

To simulate the quench from a linear to zigzag chain, the radial trapping frequencies, $\nu_{t1}(t)$ and $\nu_{t2}(t)$, decrease linearly with time at rates corresponding to the experimentally measured values. The stochastic differential equation (*) was numerically integrated between 1500 and 3000 times per quench time $\tau_Q$ in order to achieve a low statistical uncertainty.

## II.     LIFETIME OF KINKS

We investigate the stability of the observed kinks by analyzing time series of images of the ions. For each time series, after a radial quench is applied, 20 pictures with exposure times of 40 ms are taken. The disappearance of a kink typically coincides with a flip of one side of the ion chain as they reorder into a normal zigzag configuration, which happens on a time scale much shorter than the exposure time. Since kinks disappear only in connection with a change in the entire crystal structure, the primary mechanism for kink loss appears to be collisions of the trapped ions with particles from the background gas which provide enough energy to disturb the entire crystal. The lifetime of one kink is estimated from the number of pictures in which it appears. A histogram showing the number of kinks surviving after a given observation time is shown in Supplementary Figure 3. Fitting an exponential function to this histogram reveals that the kinks decay with time constant of $\tau_{\text{decay}} = 1.6$ s.

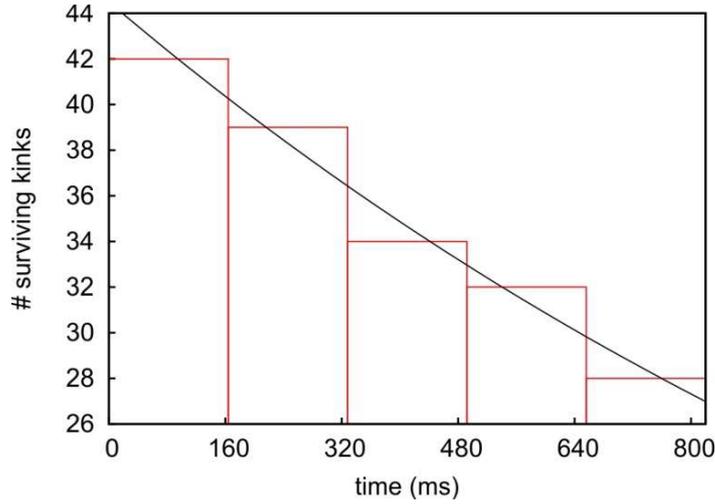

**Supplementary Figure 4 | Measurement of kink lifetime.** The histogram shows the number of kinks still present after time t (red boxes). The fit (black line) is an exponential function $f(x) = a\, e^{-t/\tau}$ with $a = 44 \pm 1$ and time constant $\tau = (1.6 \pm 0.1)$ s.

## III.     THE INHOMOGENEOUS KIBBLE-ZUREK MECHANISM

The axial confinement in an ion chain makes the inter-ion spacing spatially dependent, $a = a(x)$. Using the local density approximation, away from the chain edges the linear density $n(x) = a(x)^{-1}$ is well approximated by the inverted parabola [6,7]

$$n(x) = \frac{3\,N}{4\,L}\left(1 - \frac{x^2}{L^2}\right),$$

where $N$ is the number of ions, $L$ is half the length of the chain and $x$ the distance from the center. This leads to a spatial modulation of the critical frequency along the chain,

$$\nu_{t,c}(x) = \left[\frac{7}{2}\zeta(3)\right]^{1/2} \omega(x) \simeq 2\omega(x),$$

where $\zeta(x)$ is the Riemann zeta function. The characteristic Coulomb frequency for ions (of mass $m$ and charge $Q$) is given by $\omega(x) = \sqrt{\frac{Q^2}{ma(x)^3}}$, and ultimately makes the linear to zigzag transition inhomogeneous.

In the thermodynamic limit, the system obeys an effective time-dependent Ginzburg-Landau equation where the difference $\nu_t^2 - \nu_{t,c}^2$ governs the transition from the high-symmetry phase to the broken symmetry phase [6,7]. Consider a quench of the transverse trap frequency $\nu_t$ in the time interval $[-\tau_Q, \tau_Q]$, such that

$$\nu_t(t)^2 = \nu_{t,c}(0)^2 - \frac{\delta_0^2}{\tau_Q} t.$$

Around the critical point the transverse frequency can be linearized, $\nu_t \simeq \nu_{t,c}(0) - \delta\, t/\tau_Q$ with $\delta = \delta_0^2/[2\nu_{t,c}(0)]$. Under such a quench, as a result of the spatial dependence of $\nu_{t,c}(x)$, the zigzag phase is not formed everywhere at once but arises first in the center of the chain. In this scenario the standard KZM [1-4,32,33] breaks down and to account for the formation of kinks the mechanism must be extended to the inhomogeneous scenarios [5-8], also discussed in earlier works [13,14]. The reduced squared-frequency,

$$\epsilon(x,t) = \frac{\nu_t(t)^2 - \nu_{t,c}(x)^2}{\nu_{t,c}(x)^2},$$

governs the divergence of the correlation length and the relaxation time at the critical point

$$\xi = \frac{\xi_0}{|\epsilon(x,t)|^{1/2}}, \quad \tau = \frac{\tau_0}{|\epsilon(x,t)|^{1/2}},$$

where $\xi_0 = a = a(0)$ is set by the inter-ion spacing and $\tau_0 = \omega(0)^{-1}$ by the inverse of the characteristic Coulomb frequency. We have assumed that the system is underdamped which is the case whenever the dissipation strength $\eta$ induced by laser cooling satisfies $\eta^3 \ll \delta_0^2/\tau_Q$. The critical exponents are $\nu = 1/2$, and $z = 1$ in this regime [6,34].

The front crossing the transition satisfies $\varepsilon(x,t) = 0$ and reaches $x$ at time

$$t_F = \tau_Q \left(1 - \frac{\nu_{t,c}(x)^2}{\nu_{t,c}(0)^2}\right).$$

Relative to this time, it is possible to rewrite the reduced squared-frequency as

$$\epsilon(x,t) = \frac{t_F - t}{\tau_Q(x)},$$

in terms of the local quench rate

$$\tau_Q(x) = \tau_Q \frac{\nu_{t,c}(x)^2}{\nu_{t,c}(0)^2} = \tau_Q(1 - X^2)^3,$$

where $X = x/L$. The front velocity reads

$$v_F \sim \frac{\delta_0^2}{\tau_Q} \left|\frac{d\,\nu_{t,c}(x)^2}{dx}\right|_{x_F}^{-1} = \frac{L\,\delta_0^2}{6\,\nu_{t,c}(0)^2\,\tau_Q} \frac{1}{|X|(1-X^2)^2}.$$

The essence of the Inhomogeneous Kibble-Zurek mechanism (IKZM) is the interplay between the velocity of the front and the sound velocity. In references [5,8], the relevant velocity of perturbations is upper bounded by the ratio of the local correlation length $\hat{\xi}(x) = \xi_0 \left[\frac{\tau_Q(x)}{\tau_0}\right]^{\frac{\nu}{1+\nu z}}$ and the relaxation time scale $\hat{\tau}(x) = \hat{t}(x) = \left[\tau_0 \tau_Q(x)^{\nu z}\right]^{\frac{\nu}{1+\nu z}}$ at the freeze-out time $\hat{t}(x)$, that is, by

$$\hat{v}(x) = \frac{\hat{\xi}(x)}{\hat{\tau}(x)} = \frac{\xi_0}{\tau_0}\left[\frac{\tau_0}{\tau_Q(x)}\right]^{\frac{\nu(z-1)}{1+\nu z}} = a\omega(0).$$

The last equality holds whenever the dynamic critical exponent $z = 1$, as in an underdamped ion chain.

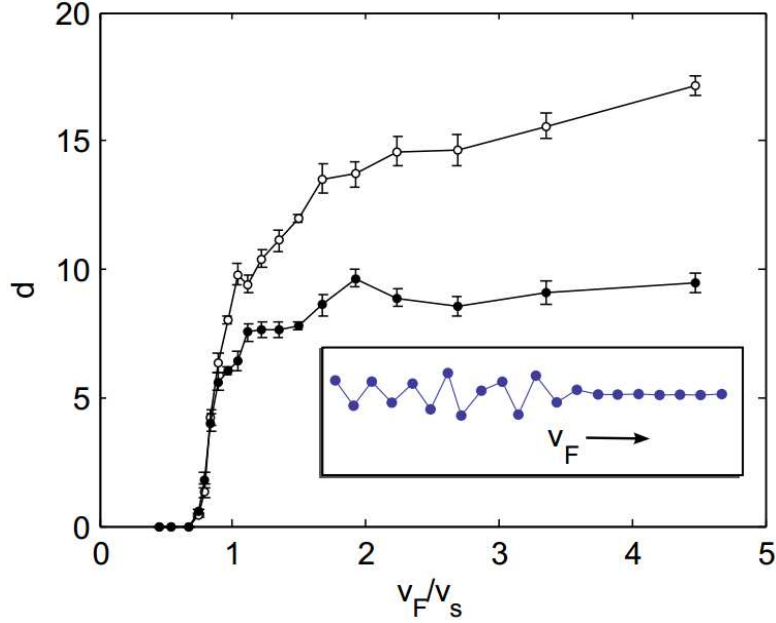

**Figure 5. Determination of the critical front velocity for kink formation.** A homogeneous ion chain is considered, in which the transverse frequency is quenched locally in a sudden way from an initial $v_t^{(i)}$ to a final $v_t^{(f)}$ frequency. This local quench propagates through the chain with a front velocity, $v_F$. The average number of kinks, $d$, as a function of front velocity, $v_F$, is measured straight after the whole chain undergoes the phase transition (open circles) and at a later time during which kinks interact and annihilate (filled circles). The onset of adiabaticity occurs for $\frac{v_F}{\hat{v}} \lesssim 0.7$. The simulations were performed for 260 Yb$^+$ ions with inter-ion spacing of 10.0 $\mu$m, and the values of the initial and final trapping potential, friction coefficient and temperature were set to $v_t^{(i)}/(2\pi) = 880$ kHz, $v_t^{(f)}/(2\pi) = 250$ kHz, $\eta = 1.3 \times 10^{-20}$ kg s$^{-1}$ and $T = 4$ mK. The time allowed for annihilation of kinks was 100 $\mu$s.

In the IKZM, the condition for kink formation to be possible is given by the inequality $v_F(x) > \hat{v}$, while adiabatic dynamics is expected otherwise. We have tested the sharpness of this inequality in Figure 5, which agrees qualitatively with similar studies in other systems [14,35]. The simulation corresponds to a homogeneous ion chain with periodic boundary conditions in which the transverse frequency is locally quenched between an initial and a final value in a sudden fashion. As a result, the velocity of the front is independent of the quench rate and the position, $v_F(x) = v_F$. There is a steep rise of the density of defects starting around $v_F \sim \hat{v}$ which signals the breakdown of adiabaticity. Further, after relaxation, the density of defects approaches a step function with respect to $v_F/\hat{v}$. This justifies the following derivation of the power-law IKZM scaling. Kink formation is expected where $v_F$ diverges, which happens for the harmonically trapped chain in the central region around $x = 0$ where the density is approximately uniform. The velocity of the front decreases at larger values of $x$. Given that the axial density $n(x)$ is well-described by an inverted parabola, defects can be formed only around the center of the cloud, in the interval $[-\hat{x}, \hat{x}]$. Generally $\hat{x}$ is to be found numerically. However, if the defect formation is restricted to a region $\hat{x} \ll L$, then one can set

$$\hat{x} \approx \frac{L^2 \delta_0^2}{6 \, v_{t,c}(0)^2 \, \tau_Q \hat{v}}.$$

Under the assumption $\hat{x} \ll L$ one finds the estimate[6,7]

$$d = \frac{2\hat{x}}{\hat{\xi}} = \frac{L^2 \delta_0^2}{3 \, v_{t,c}(0)^2 \, \hat{v}^2} \left(\frac{\delta_0^2}{\tau_Q}\right)^{4/3}.$$

Note that setting $\tau_Q(x) = \tau_Q$ and $\hat{\xi}(x) = \hat{\xi}(0)$ is consistent with $\hat{x} \ll L$. We note that without restricting to this limit, there is no reason to expect a power-law scaling, see [8].

## IV. DOUBLING OF THE INHOMOGENEOUS KZM SCALING

We have seen that provided that $\hat{x} \ll L$ it is possible to observe the IKZM power-law scaling. Now we shall consider the special case in which $\hat{x} \ll L$ holds, but $2\hat{x} \sim \hat{\xi}$, so that typically one obtains 0 or 1 defects per realization. In the analogous homogeneous scenario, $2L \sim \hat{\xi}$, the doubling of the scaling was experimentally demonstrated in [36], and further derived by three independent studies in a variety of systems [15-17]. Here, we follow [17] to predict a doubling of the IKZM. Letting $y$ be the transverse direction of the ions, we recall the definition of the order parameter in the continuous limit [6,7] $(-1)^n y_n \to \psi(x)$. We introduce the total local density of defects [37], which in terms of the sign function is given by $n(x) = \partial_x \, sgn[\psi(x)]/2$. For a chain with a kink the integral $q = \int n(x) \, dx = +1$ while in the presence of a single anti-kink is $-1$. Averaging over many realizations the sum of topological charges (or difference in the number) of kinks and antikinks vanishes,

$$\langle n \rangle = \int_{-\hat{x}}^{\hat{x}} \langle n(x) \rangle dx = 0,$$

so we consider the average mean square

$$\langle n^2 \rangle = \int_{-\hat{x}}^{\hat{x}} \langle n(x)n(x') \rangle dx dx'.$$

We note that if $2\hat{x} \sim \hat{\xi}$, then $\langle n^2 \rangle \approx p_{1kink}$, the probability of observing a single defect (either kink or anti-kink). We introduce the density-density correlation function $g(x,x') = \langle n(x)n(x') \rangle$ and assume, that for $|x - x'| \sim \hat{x} \ll L$, $g(x,x') = g(x - x', 2\hat{x})$. That is, for small displacements of the argument $x$ in which the amplitude of the order parameter $\psi(x)$ (width of the chain) is approximately constant, $g(x,x')$ is assumed to be translationally invariant. Moreover, we note that from dimensional analysis, the short distance pair-correlation function relevant to the case $2\hat{x} \ll \hat{\xi}$ behaves as

$$g(x, 2\hat{x}) = \frac{1}{\hat{\xi}^2} \bar{g}\left(\frac{x}{\hat{\xi}}, \frac{2\hat{x}}{\hat{\xi}}\right).$$

The probability to observe a single defect (either kink or antikink) can be estimated to be

$$p_{1kink} = \frac{2\hat{x}}{\hat{\xi}} \int_{-\hat{x}/\hat{\xi}}^{\hat{x}/\hat{\xi}} \bar{g}\left(y, \frac{2\hat{x}}{\hat{\xi}}\right) dy \approx \frac{1}{2}\left(\frac{2\hat{x}}{\hat{\xi}}\right)^2 \bar{g}\left(0, \frac{2\hat{x}}{\hat{\xi}}\right).$$

We expect $\bar{g}\left(0, \frac{2\hat{x}}{\hat{\xi}}\right)$ to be approximately constant and of order unity for different $\hat{\xi}$, so that

$$p_{1kink} \propto \frac{L^4 \delta_0^4}{v_{t,c}(0)^4 \hat{v}^4} \left(\frac{\delta_0^2}{\tau_Q}\right)^{8/3}.$$

## V. OVERVIEW OF EXPERIMENTS RELATED TO KZM

Experimental data are consistent with KZM, but, so far, did not convincingly test its central prediction – scaling of the number of created defects with the rate of quench. Thus, defects were seen in the wake of phase transitions in liquid crystals [38,39], $^3$He [40,41] and superconductors [42] where it was also confirmed [43] that they are anticorrelated as predicted by KZM. Observations consistent with KZM were made in other systems [44-45], although initial observation of post-transition vortices in $^4$He (see [46]) was later attributed to inadvertent stirring [47], and attempts to see quench-generated defects in $^4$He were so far unsuccessful. Recent beautiful experiments in Bose-Einstein condensates [48,49] confirm defect production, but not the scaling. The most interesting recent development is the observation [19,50] of KZM scaling in multiferroics. Interpretation in terms of KZM is suggestive, but is based on theoretical predictions that have not been confirmed by measurement of equilibrium critical exponents. Moreover, scaling breaks down for very fast quenches. Similarly, in tunnel Josephson junctions, scaling of the probability to trap a single flux line has been measured [17,18,51], but with exponents that require additional assumptions (about e.g., external fields) in order to be consistent with KZM.

The main difficulty with observing KZM scaling is the fractional dependence of defect densities in homogeneous systems. This means that one has to change quench time over orders of magnitude for a reliable signal. A way to enhance scaling of defect density on quench rate that relies on causality (central to KZM) was suggested [5] in investigation of systems in harmonic traps, where symmetry breaking starts in the center of the trap, and the transition front propagates with the speed set by the quench rate and the local gradient of the potential. Where the front is faster than the speed of the relevant sound, different parts of the condensate cannot communicate, and break symmetry independently as would be the case for the homogeneous transition [13,14]. However, further from the trap center, the phase front is slower, and as soon as it is slower than the sound, defect production ceases (as the outlying parts of the system inherit broken symmetry from their neighbours). Thus, in harmonic traps quench rate controls both the density of defects in the homogeneous region and the size of the fraction of the trap where they can be produced. This leads to steeper scaling [5,6,8]. In our case we expect scaling that is (i) enhanced by inhomogeneity of the harmonic trap and (ii) doubled by the focus on the probability of finding a single defect (rather than counting many defects) as was the case in [18,51].